\documentclass[twocolumn,showpacs,preprintnumbers,amsmath,amssymb]{revtex4}

\usepackage{graphicx}
\usepackage{dcolumn}
\usepackage{bm}

\begin{document}

\title{Asymmetric superradiant scattering and abnormal mode amplification induced by atomic density distortion}

\author{Zhongkai Wang, Linxiao Niu, Peng Zhang, Mingxuan Wen, Zhen Fang, Xuzong Chen, and Xiaoji Zhou}\thanks{Electronic address: xjzhou@pku.edu.cn}

\affiliation{School of Electronics Engineering $\&$ Computer
Science, Peking University, Beijing 100871, China}

\date{\today}

\begin{abstract}
The superradiant Rayleigh scattering using a pump laser incident
along the short axis of a Bose-Einstein condensate with a density
distortion is studied, where the distortion is formed by shocking
the condensate utilizing the residual magnetic force after the
switching-off of the trapping potential. We find that very small
variation of the atomic density distribution would induce remarkable
asymmetrically populated scattering modes by the matter-wave
superradiance with long time pulse. The optical field in the diluter
region of the atomic cloud is more greatly amplified, which is not
an ordinary mode amplification with the previous cognition. Our
numerical simulations with the density envelop distortion are
consistent with the experimental results. This supplies a useful
method to reflect the geometric symmetries of the atomic density
profile by the superradiance scattering.
\end{abstract}
\pacs{03.75.Kk, 42.50.Gy, 42.50.Ct, 32.80.Qk}.

\maketitle
\section{Introduction}

Matter-wave superradiance scattering is induced by the highly
anisotropic geometry of Bose-Einstein Condensate (BEC). Ever since
the first observation of
superradiant scattering from the BEC, a series of experiments~\cite%
{Inouye1999science,Schneble2003scince,1999,decay3,Yutaka,decay1,yang,deng,decay4}
have been explored to simulate the origin of this phenomenon and the
dynamical properties of the light-atom coupling process. Experiments
of different geometries, such as the Rayleigh and Raman superradiant
scattering, in which the lights are pumped in the axial or radial
direction~\cite{end1,end2} were intensively studied. The fledged
quantum theories describing the establishment of superradiant by
mode competition~\cite{Bar-Gill2007,zhou,Moore1999prl,Pu2003prl}, or
semi-classical theories~\cite{Zobay2006pra} describing the gain
process are given and checked to be consistent with the relevant
experiments in pretty accuracy. Because of the high controllability
of matter-wave systems, the atomic superradiance has found its
applications in areas such
as coherent imaging, quantum memory and manipulation~\cite%
{decay1,application,memory}.

The spatial anisotropy is crucial for the generation of
superradiance. In this process the photons are emitted into the
end-fire modes.
 In previous works about superradiant radiation of BEC, the phases and densities of the condensates can
be treated as uniformly distributed, if we neglect the boundary effects. Before switching off the trap, by
tuning the incident intensity, detuning, and angle
of the pumping beam, various kinds of superradiant processes of their own
features have been realized \cite{Inouye1999science, Schneble2003scince,zhou,Lu}.
However, besides the experimental configuration, the local properties such
as the local phase or density distribution are also very important factors
in determining the dynamics of light-atom coupling systems, resulting in
interesting phenomena in matter-wave superradiant processes. Unfortunately, the relevant
research of superradiance in this area has not been paid enough attention as it deserved.

In this paper, we investigate the superradiant Rayleigh scattering
by a pump laser incident along the short axis of a density distorted
atomic condensate confined in an unbalanced magnetic trap. Even
though, in our experiment, the density distortion of the condensate
is tiny in a time-of-flight imaging, it induces rather prominent
asymmetry in the atomic population for the first-order superradiant
scattering modes. Unlike the ordinary optical mode amplification,
the superradiance scattering optical field in the diluter region of
the condensate is better amplified than that in the denser region.
Our numerical simulation using the coupled Maxwell-Schr\"{o}dinger
equation with a density envelope modulation shows a good consistence
with the experimental results. Thus the matter-wave superradiance
supplies a useful tool for monitoring the spatial symmetry of the
density distribution of BECs, without any too complicated optical
laser techniques being involved.

The paper is arranged as follows. Section 2 gives our experimental description and results of asymmetrically populated
scattering modes of the density distorted matter-wave superradiance. In Section 3, we present the theoretical model describing the superradiance of the distorted condensate. The effects of the atomic density distortion on the scattering process are analyzed and numerically simulated in Section 4. Finally, the conclusion is summarized in Section 5.

\section{Experimental description}

\begin{figure}
\begin{center}
\scalebox{0.40}[0.38] {\includegraphics*
[5pt,250pt][590pt,830pt]{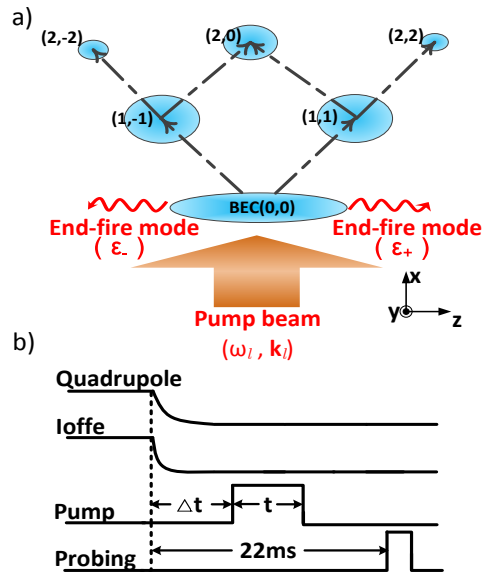}}
\end{center}
\caption{(a) An elongated condensate is pumped by a single
off-resonant laser beam $(\protect\omega _{l},\mathbf{k}_{l})$. The
beam is collectively scattered to two end-fire modes and the atoms
get recoiled to the discrete side-modes $(m,n)$. (b) The time
sequence of our experiment. When the magnetic trap is turned off,
there is a residual magnetic fields generated, resulted from the
different decay rates of the quadrupole and Ioffe trap. Then after a
delay $\Delta t$, a pump field with duration $t$ and intensity $I$
is incident upon the BEC along the short axis of the condensate to
initiate the superradiant scattering. Finally, after a 22ms
time-of-flight, a probing beam is applied to the system to get the
absorption images of the scattered side-modes. } \label{sequence}
\end{figure}

For a BEC of highly anisotropic geometry as shown in
Fig.~\ref{sequence}(a), when it is exposed to a laser beam with wave
vector $\mathbf{k}_{l}$ (and frequency $\omega _{l}$) along its
short axis (the $\mathbf{\hat{x}}$ direction), as the typical
superradiance scattering demonstrated~\cite{Inouye1999science,
Schneble2003scince}, it will scatter photons to the end-fire modes
with wave vector $\{\mathbf{k}_{nm}\}$ along the longitudinal (the
$\mathbf{\hat{z}}$ direction) axis of the condensate. At the same
time, the atoms within the condensate will get recoiled to the
discrete side-modes ($n,m$) with momentum $n\hbar
\mathbf{k}_{l}+m\hbar \mathbf{k}_{nm}$ ($n$, $m$ are integer). Here,
the frequency of the optical fields generated in each end-fire mode
shall be different with one another for energy-momentum
conservation, depending on the status of the BEC and the
experimental configuration. However, one can approximate that
${k}_{nm}$ $\approx $ ${k}_{l}$ $\approx $ ${k}$ because of the
dispersion relation of light. After the superradiant process has
been initiated spontaneously, the recoiled atoms would interfere
with the condensate, forming a matter-wave grating, and then be
amplified by stimulated Rayleigh scattering. At the initial stage,
the population of atoms in each recoil side-mode ($n,m$) grows
exponentially following the gain equation ~\cite{Inouye1999science},

\begin{equation}
\dot{N}_{n,m}=(G_{n,m}-\Gamma _{n,m})N_{n,m},  \label{gain}
\end{equation}%
where $N_{n,m}$ is the atomic number of side-mode (n,m), $G_{n,m}$ and $%
\Gamma _{n,m}$ are the gain and loss coefficient, respectively,
describing the amplification and decoherence of the process.

In our experiment, by laser cooling and rf-evaporation cooling, a
cigar-shaped BEC with $N=1.4\times 10^{5}$ $^{87}$Rb atoms is
prepared in the $ |F=2,m_{F}=2\rangle $ hyperfine ground state with
a long axis $ L=80\mu m$ and short axis $D=8\mu m$ in an
Ioffe-quadrupole trap with trapping frequencies $\omega_x=\omega_y=2
\pi \times220Hz$ and $\omega_z=2 \pi \times20Hz$. In order to
manipulate the atomic density, we shock the condensate by generating
a magnetic force via abruptly switching off the trapping potential
by the fact that the residual currents remained in the coils of the
quadrupole and Ioffe trap decay at different rates. The time control
sequence is shown in Fig.\ref{sequence}(b). This results in a BEC
with a density profile like that in Fig.~\ref{Fig.2}(a) measured by
absorption imaging after free expansion. It can be seen that the
density of the BEC is a little higher on the left side. Then after a
delay $\Delta t=500\mu s$, when the trapping potential has vanished,
a laser field with tunable pulse length $t$ and wave length $\lambda
_{l}=780nm$, which is red detuned by $\delta =-0.88GHz$ to the
$|F=2,m_{F}=2\rangle$$\rightarrow$$|F'=3,m_{F}'=3\rangle$
transition, is incident on the condensate along the radial direction
to initialize the matter-wave superradiance. The diameter of the
pump laser is about $2.2mm$, which is more than $20$ times the
longitudinal full width at half maximum of the condensate, thus we
can treat the laser field as plane wave. Finally, after a $22ms$
time-of-flight(TOF) for free expansion, the momentum distribution of
the matter-wave is measured by absorption imaging.

The Rayleigh scattering typically results in fan-shaped patterns
with our pump pulse intensity $I=80mW/cm^{2}$ and the pulse time $t
=100 \mu s$. The atomic populations in two first-order forward
scattering modes ($1,-1$) and ($1,1$) are measured with the pump
beam incident to a BEC before and after the switching-off of the
trapping potential, respectively, as shown in the TOF images of the
condensate in Figs.~\ref{Fig.2}(b) and ~\ref{Fig.2}(c). In the case
that the superradiance takes place before the magnetic trap is shut
down, the density of a dilute BEC (the interatomic interaction can
be neglected in the dilute BEC) is symmetrically distributed within
the trap, because the size of the BEC is small compared to the
character length scale of the spacial fluctuations of the trapping
potential, resulting in a density profile of the Gaussian form
symmetric about its geometric center. It is also true that envelop
of the BEC should be symmetric even if the condensate is dense. In
this situation, the wave-function of the ground state shall take the
Thomas-Fermi form in mean-field approximation because of the
presence of the atomic interactions. The superradiance of such
symmetric BEC has been studied
\cite{Inouye1999science,Schneble2003scince,1999,decay3,Yutaka,decay1,yang,deng,decay4};
however, research of that for the asymmetric BEC as that shown in
Fig.~\ref{Fig.2}(a) has been lack. Compared to the matter-wave
superradiance for the symmetric BEC with equal atomic populations in
both the two first-order side-modes as shown in Fig.~\ref{Fig.2}(b),
it can be seen in Fig.~\ref{Fig.2}(c) that, there is an asymmetric
atomic occupation in these two modes; the rightward end-fire mode is
unexpectedly more drastically amplified than the leftward end-fire
mode, even though the density of the atomic cloud which functions as
the gain medium for the light amplification is lower on the right
half of the condensate. The rightward end-fire mode scatters the
atoms into the ($1,-1$) side-mode. This results in a higher atomic
population in this mode than that in the ($1,1$) mode. To get
insight in this unwonted mode amplification, we have to go into the
details of the light-atom interaction.
\section{ Theoretical model of the density distorted matter-wave Superradiance}

In the electric dipole approximation, the coupling between the
superradiant matter-wave and light fields (the pump field and the
end-fire modes) can be
described by semiclassical Maxwell-Schr\"{o}dinger equations~\cite%
{Zobay2006pra},
\begin{eqnarray}
i\hbar \frac{\partial }{\partial t}\psi\left( \mathbf{r},t\right)
&=& -\frac{\hbar ^{2}}{2M}\Delta \psi\left( \mathbf{r},t\right)\nonumber\\
&+&\frac{(\mathbf{d}\cdot\! \mathbf{E}^{(+)})(\mathbf{d}%
\cdot \mathbf{E}^{( -)})}{ \hbar \delta} \psi\left( \mathbf{r}%
,t\right),  \label{M-S1} \\
\frac{\partial ^{2}\mathbf{E}^{(\pm )}}{\partial
t^{2}}&=&c^{2}\Delta
\mathbf{E}^{(\pm )} -\frac{1}{\varepsilon_{0}}\frac{\partial ^{2}\mathbf{P}%
^{(\pm )}}{\partial t^{2}},  \label{M-S2}
\end{eqnarray}
where, $\psi \left( \mathbf{r},t\right) $ is the atomic wave function. $%
\mathbf{E}^{(\pm )}$ are the positive and negative frequency parts
of the
electric field. $\mathbf{P}^{(\pm)}=-\mathbf{d}|\psi (\mathbf{r},t)|^{2}%
\frac{\mathbf{d}\cdot \mathbf{E}^{(\pm)}(\mathbf{x},t)}{h\delta }$
is the density distribution of the atomic electric dipoles. $M$ is
the atomic mass, $\mathbf{d}$ the atomic dipole moment operator,
$\varepsilon _{0}$ the vacuum dielectric constant, and $c$ the speed
of light in vacuum.\ Here, the local density distribution of the
condensate $|\psi (\mathbf{r},t)|^{2}$ has to be taken into
consideration in Eq.~(\ref{M-S2}) via $\mathbf{P}^{(\pm)}$, because
the modification of the light fields propagating through the
matter-wave grating is prominent in the conditions of the present
experimental parameters. In our experiment, the atomic gas is so
dilute that the influence from the inter-particle interaction upon
the atomic motion is very weak. Therefore, the nonlinear matter-wave
self interacting term is not included in this equation, and the
atomic loss term resulted from the nonlinear effects in the time
scales of our experiment should also be
negligible~\cite{decay3,decay1,decay4}.

\begin{figure}[tbp]
\begin{center}
\includegraphics[width=8cm]{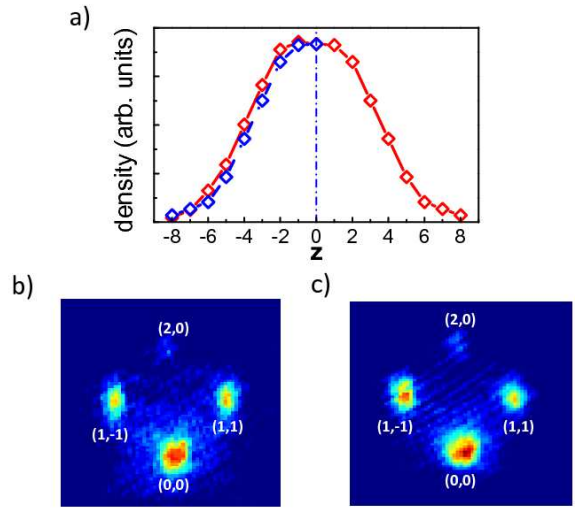}
\end{center}
\caption{(a) The experimentally measured atomic number distribution
(in arbitrary unit) along the axis of the condensate, obtained by
using the time-of-flight absorption imaging method. The blue
dash-dotted curve is the mirror image of the atomic distribution of
the right half of the BEC, implying a higher atomic density on the
left side. Absorption images of the condensate after the matter-wave
superradiance process are presented in (b) for symmetric BEC, and
(c) for distorted BEC in which the experimental data is chosen as
$\Delta t=500\protect\mu s$; both images are obtained by absorption
imaging after 22ms of free expansion and averaged over four
experiments at $t =100\protect\mu s$.} \label{Fig.2}
\end{figure}

The Fresnel number $F=\pi D^{2}/(4\lambda _{l}L)$ of the optical
pump field in our experiment is around $1$, so that we can using the
quasi-2D model to investigate the dynamics of the condensate in the
superradiant processes. In our experiment, the atomic gas is dilute,
thus one can decompose the atomic wave function in the discrete
modes~\cite{Zobay2006pra},
\begin{eqnarray}
\psi \left( \mathbf{r},\mathbf{t}\right) &=&\sum_{n,m}\frac{\psi _{n,m}(z,t)%
}{\sqrt{A}}e^{-i(\omega _{n,m}t-nkx-mkz)},
\end{eqnarray}
where $\psi _{n,m}$ represents the wave function of the side-mode
($n,m$), and $\hbar\omega _{nm}=(n^{2}+m^{2})\hbar\omega _{r}$ is
the kinetic energy of one single atom in this mode with $\omega
_{r}=\frac{\hbar k^{2}}{2M}$ the single photon recoil frequency. $A$
is the effective cross section area of the condensate. In the case
of an symmetric BEC, $A$ should be constant. Atomic number in the
(n,m) mode can be calculated as
\begin{eqnarray}
N_{n,m}=\int|\psi_{n,m}|^2\mathrm{d}z.
\end{eqnarray}
Similarly to the discrete expansion of the matter-wave function, the
optical fields can be expressed in the form of running wave with few
frequency components, because the band width of the laser beam is
very narrow,
\begin{eqnarray}
E^{(\pm )}(\mathbf{r},t) &=&\varepsilon _{l}\mathbf{e}_{y}e^{\pm
i(\omega _{l}t-kx)}/2\nonumber\\
&+&\varepsilon _{+}(z,t)\mathbf{e}_{y}e^{\pm
i(\omega _{+}t-kz)} \nonumber\\
&+&\varepsilon _{-}(z,t)\mathbf{e}_{y}e^{\pm
i(\omega _{-}t+kz)},
\end{eqnarray}
where $\varepsilon _{l},\varepsilon _{\pm }$ are the envelope
functions of the pump laser and the end-fire modes, respectively
($+$ corresponds to the end-fire mode propagating along the positive
$z$ direction, and $-$ represents that in the opposite direction).
$\mathbf{e}_{y}$ is the unit polarization vector of the electric
field along the $y$ direction.

Under the slowly-varying-envelope approximation (SVEA) and using the
transformation $\psi _{n,m}\rightarrow \frac{\psi _{n,m}\sqrt{k}}{\sqrt{A}}$%
, $\varepsilon _{\pm ,l}\rightarrow e_{\pm ,l}\sqrt{\frac{\hbar \omega k}{%
2\epsilon _{0}A}}$, $\tau =2\omega _{r}t$, and $\xi =kz$, Eq.
(\ref{M-S1}) can be recast to the following dimensionless form,
\begin{eqnarray}
i\frac{\partial \psi _{n,m}(\xi ,\tau )}{\partial \tau }
&=&-\frac{1}{2}\frac{\partial ^{2}\psi _{n,m}(\xi ,\tau )}{\partial
\xi ^{2}}
-im\frac{\partial \psi _{n,m}(\xi ,\tau )}{\partial \xi }  \nonumber \\
&+&\kappa \lbrack e_{+}^{\ast }(\xi ,\tau )\psi _{n-1,m+1}(\xi ,\tau
)e^{i(n-m-2)\tau } \nonumber\\
&+&e_{-}^{\ast }(\xi ,\tau )\psi _{n-1,m-1}(\xi
,\tau )e^{i(n+m-2)\tau }
\nonumber \\
&+&e_{+}(\xi ,\tau )\psi _{n+1,m-1}(\xi ,\tau )e^{-i(n-m)\tau }\nonumber\\
&+&e_{-}(\xi ,\tau )\psi _{n+1,m+1}(\xi ,\tau )e^{-i(n+m)\tau }],
\label{Semi}
\end{eqnarray}
where $\kappa =\frac{g}{2\omega _{r}}\sqrt{kL}$ is the dimensionless
coupling constant with $g=\frac{|\mathbf{d}|^{2}\varepsilon
_{l}}{2\hbar ^{2}\delta }\sqrt{\frac{\hbar \omega_{l}} {2\epsilon
_{0}AL}}$.

Neglecting the retardation effects, the envelope functions $e_{\pm
}$ of the end-fire modes in the dimensionless form are given by
\begin{eqnarray}
e_{\pm }(\xi ,\tau ) &=&-i\frac{\kappa }{\chi }\int \mathrm{d}\xi
^{\prime }\sum_{n,m}e^{i(n\pm m)\tau } \nonumber\\
&\times& \psi _{nm}(\xi
^{\prime },\tau)\psi _{n+1,m\mp 1}^{\ast }(\xi ^{\prime },\tau),
\label{e}
\end{eqnarray}%
with $\chi =\frac{ck}{2\omega _{r}}$. The two end-fire modes are
propagating
against each other, hence the spacial integration for $e_{+}$ is taken from $%
-\infty $ to $\xi $, while that for $e_{-}$ is taken from $\xi $ to
$\infty $.

There is one thing that is important we have to emphasize. Besides
the fact that, during the interval between the turning off of the
magnetic trap and switching-on of the pump pulse, the density
distribution of the condensate is distorted by the residual magnetic
field as shown in the experimental data in Fig.~\ref{Fig.2}(a), the
condensate is also accelerated by the unbalanced residual magnetic
force. This has been confirmed by the measured center of mass motion
of the condensate in our experiment. Then after long enough time, at
which the magnetic force died out, the BEC acquires a constant
velocity $v_{0}$. In a typical matter-wave interference experiment,
the interference patterns of the scattered matter-wave is
sensitively dependent on the local phase and density distributions
of the condensate. However, we find that, in our situation, the
local phase imprinted upon the condensate by the atomic acceleration
has little influence upon the dynamics of the condensate (for making
good consistency of the different parts of the paper, the detailed
analysis of the influence upon the atomic motion from the
acceleration is presented in the Appendix). In drastic contrast,
even a tiny disturbance of atomic density distribution would
generate prominent asymmetric interference patterns of the scattered
matter-wave in the superradiant process. This makes our experiment a
rather robust technique for measuring the geometric symmetry of BECs
with high precision.

\section{The density modulation analysis and numerical simulation}

\begin{figure*}[tbp]
\begin{center}
\includegraphics[bb=110 324 554 585,width=12cm]{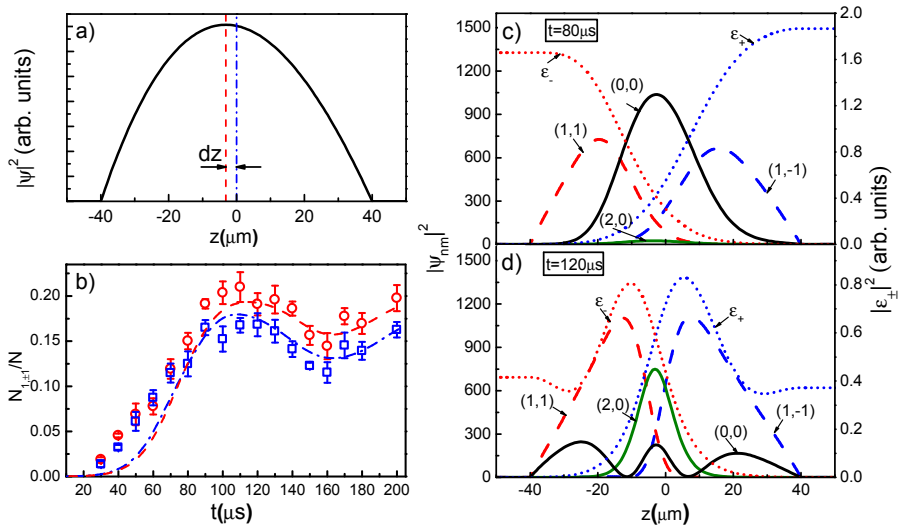}
\end{center}
\caption{(a) The profile of the wave function used in our
simulation. It is a linearly distorted wave function in Thomas-Fermi
approximation. $dz$ measures the displacement of the peak density
point of the condensate from the geometric center. (b)
Experimentally measured atomic populations of (1,1) (red circle) and
(1,-1) (blue square) side-modes in comparison to the simulated
results for (1,1) (red dashed line) and (1,-1) (blue dash-dotted
line), in the matter-wave superradiant experiment. The light-atom
coupling constant is chosen $g=3.6\times 10^{6}$ $s^{-1}$ according
to the experiment, with the fitted parameter $l=-500\mu m$. The
numerical results at $t =80\protect\mu s$ and $t =120\protect\mu s$
are presented in (c) and (d), respectively. The density profile
$|\psi_{0,0}|^{2}(z)$ of the $\left( 0,0\right) $ mode is in black
solid curve, $|\psi_{1,-1}|^{2}(z)$ in blue dashed curve,
$|\psi_{1,1}|^{2}(z)$ in red dashed curve and $|\psi_{2,0}|^{2}(z)$
in green solid curve; the strength of the end-fire mode
$\protect\varepsilon _{+}$ is in blue dotted curve, and
$\protect\varepsilon _{-}$ is in\ red dotted curve.} \label{simu}
\end{figure*}

Now we consider the influence upon the atomic motion and modes
amplification from the distorted density distribution of the
condensate. As show in the experimental data in Fig.~\ref{Fig.2}(a),
the left half of the condensate has a higher density than that of
the right half. Thus the superradiant scattering in the denser side
would have a greater gain leading to a coordinate-dependent coupling
constant $\kappa$ in Eq.~(\ref{Semi}) via the $z$-dependence of the
effective cross section area of the condensate $A$. To describe such
a mechanism that the atomic density distortion modifies the
matter-wave superradiance, we multiply a factor $S^{-2}(z)$ to the
effective cross section area of the condensate $A\rightarrow
A/S^{2}(z)$. In the regime where the distortion is very weak, $S(z)$
can be approximated up to the linear order of $z$,
$S(z)=(1+\frac{z}{l}+O(z^{2}))$, with a parameter $l$ that should be
fitted from the experimental data. During the radial expansion, the
atomic number $N(z)=|\psi_{00}(z)|^2$ will not change, thus the
asymmetric cross section $A/S^{2}(z)$ would lead to a transformation
of the coupling constant $\kappa\rightarrow \kappa(1+\xi /\Lambda )$
with $\Lambda =kl$. The envelope of the density distribution of the
initial distorted condensate $\psi$ used in our simulation is shown
in Fig.~\ref{simu}(a). It is a modified function in Thomas-Fermi
approximation$\psi=\sqrt{N/A}(1+\xi /\Lambda )\psi _{s}$, where
$|\psi
_{s}(z)|^2=3/4L^{3}[(L/2)^{2}-z^{2}]\Theta (L/2-|z|)$ with $%
\Theta \left( z\right) $ the Heaviside step function. $\psi_{s}$ is
a seed wave function indicating one atom in a side-mode. To measure
the asymmetry of the initial BEC, we define a parameter $dz$ in
describing the displacement of the peak density point of the
condensate from its geometric center. In our simulation, $dz=3.16\mu
m$. This tiny shift is hardly measurable in ordinary imaging system.
However, by monitoring the symmetry of the dynamics of the scattered
matter-wave in the superradiant process, one can obtain information
of the geometric symmetry of the condensate, which is exclusively
available in using the \emph{in situ} imaging
methods~\cite{insitu1,chin}.

\subsection{Early stage}
At the early stage of the superradiant process, depletion of the
condensate is negligible and only the two first-order side-modes
are dominantly excited with exponential amplification. The coupling
between each of these modes to the ground state reads,

\begin{eqnarray}
i\frac{\partial \psi _{1,\pm 1}}{\partial \tau } =-\frac{1}{2}%
\frac{\partial ^{2}\psi _{1,\pm 1}}{\partial \xi ^{2}}-i\frac{%
\partial \psi _{1,\pm 1}}{\partial \xi }+\kappa e_{\pm }^{\ast }S^{2}\psi _{0,0},
\label{psi2}
\end{eqnarray}%
with the electric field in each end-fire mode taking the following
form,
\begin{equation}
e_{\pm}(\xi,\tau)=\frac{-i\kappa S(\xi )}{\chi}\int\mathrm{d}%
\xi^{\prime }\psi _{00}(\xi ^{\prime })\psi _{1,\mp1}^{\ast }(\xi
^{\prime },\tau)S^{2},\label{e2}
\end{equation}
where, $\psi _{00}=\sqrt{N}\psi _{s}$\emph{, }$\psi _{1,\pm 1}=\sqrt{%
N_{1,\pm 1}}\psi _{s}$. The first two terms on the right hand side
of Eq. (\ref{psi2}) represent the kinetic energy of each side-mode,
while the last term describes the coupling between $\psi _{00}$ and
$\psi _{1,\pm 1}$ by the electric field $e_{\pm }$, respectively.

In the early stage $\tau \rightarrow 0$, the first two kinetic terms
in Eq. (\ref{psi2}) are negligible, leading to an exact expression
of the atomic population in each side-mode as,

\begin{equation}
N_{1,\pm 1}(\tau )=N_{1,\pm 1}(0)e^{G_{1,\pm 1}\tau },  \label{Nexp}
\end{equation}
with
\begin{equation}
G_{1,1} =\frac{N\kappa ^{2}}{{\chi }}\int_{-\infty }^{\infty
}\mathrm{d}\xi\psi _{s}^{\ast }\psi _{s}S^{2} \int_{-\infty }^{\xi
}\mathrm{d}\xi ^{\prime }\psi _{s}^{\ast }\psi _{s}S^{2}, \label{G1}
\end{equation}

\begin{equation}
G_{1,-1} =\frac{N\kappa ^{2}}{{\chi }}\int_{-\infty }^{\infty
}\mathrm{d}\xi\psi _{s}^{\ast }\psi _{s}S^{2} \int_{\xi }^{\infty
}\mathrm{d}\xi ^{\prime }\psi _{s}^{\ast }\psi _{s}S^{2}. \label{G2}
\end{equation}
The integrals in Eqs. (\ref{G1}) and (\ref{G2}) can be numerically
carried out to give in the dimensionless notion $G_{1,1}=16.7$ and
$G_{1,-1}=15.9$, implying that the atomic populations in these two
side-modes increase exponentially at the beginning of the
matter-wave superradiant process as shown in the experimental data
and numerical simulation in Fig.~\ref{simu}(b). Even though such
growing is unbalanced for the $\left( 1,\pm 1\right) $ modes, the
asymmetry of the matter-wave superradiance is hardly visible in
experiment in this limit, because the atomic populations in these
two modes are too small. However, the numerical simulation is
consistent with the experimental data in predicting the exponential
growing of the atomic numbers in the side-modes in this stage.

\subsection{Long pulse regime}

In the long pulse regime, things are quite different.
Equation~(\ref{Nexp}) is not valid in describing the dynamics of the
matter-wave superradiance any more. By solving the semiclassical
equations (\ref{Semi}) and (\ref{e}), we can calculate the atomic
populations on the\ different side-modes. Our numerical stimulation
has taken into account the\ coupling between the side-modes of high
orders to a large cut-off. The result in this regime is also
contained and shown in Fig.~\ref{simu}(b) with a fitted value $l
=-500\mu m$. To fit the experimental result with the quasi-two
dimensional model, the numerical result is scaled by a factor of
$2$. This discrepancy is partly due to the collision between the
side-modes and the condensate \cite{Bar-Gill2007,collision}. At the
same time, in a real experiment, plenty of atoms in the condensate
far away from the center with a much lower density are not included
in our theoretical model. The superradiant gain is dependent on the
density of the BEC, thus the collective gain does not occur below a
critical density~\cite{lowd}, then a big amount of these atoms will
stay in their initial state, leading to a greater atomic population
in the ($0,0$) mode in experiment than that predicted by the
numerical calculation.

It is known that, for BEC prepared in trap of perfect symmetry in
the longitudinal direction, there is no asymmetry in the atomic
occupations among the different side-modes in the time domain in the
matter-wave superradiant process, provided that the pump field is
incident on the condensate along the radial direction of BEC and the
atomic distribution maintains unperturbed during the superradiant
process. For example, for the superradiant process happens in the
trap, there is no asymmetry between the two side-modes $(1,\pm 1)$
as shown in Fig.~\ref{Fig.2}(b). While, for BEC in the unbalanced
trap with the time controlling sequence chosen such that $\Delta
t=500\mu s$ in our experiment, the condensate is shocked and
distorted by the residual magnetic force to take the shape like that
in Fig.~\ref{Fig.2}(a). The matter-wave superradiance shows
unconventional mode amplification behaviors. It is seen in
Fig.~\ref{simu}(c) that, the $\varepsilon_{+}$ and $\varepsilon_{-}$
end-fire modes are mainly amplified and concentrated in the right
part and left part of the condensate, respectively, and the
condensate here functions as the gain medium for the optical mode
amplification. However, for longer pulse durations, even though the
density in the left side of the condensate is higher than that in
the left side, the ($1,-1$) mode grows faster than the ($1,1$) mode,
implying that $\varepsilon_{+}$ end-fire mode is more quickly
amplified than that propagating in the opposite direction. This is
in contradiction with our common cognition that the optical mode
should be better amplified provided that the gain medium is denser.

The above abnormal side-mode amplification can be well explained by
investigating the mode competition in the central region of the
condensate. As the duration of the pump laser is prolonged, the
atoms in the outer regime of the condensate have been greatly
exhausted (as shown in Fig.~\ref{simu}(c)), atomic scattering
process in the central part of the condensate begins to decisively
influence the atomic occupations among the side-modes. In this
region, the atomic cloud in the left part (i.e., $z<0$) of the
condensate is denser than that in the right part, making the $e_{+}$
mode grow faster than the $e_{-}$ mode. In fact, the faster growing
of $e_{+}$ in the central region is true in the full range of the
pulse duration, including the early stage of the process, as shown
in Figs.~\ref{simu}(b) and ~\ref{simu}(c). This makes the wave
function of the ($ 1,-1 $) mode have a fatter tail than that of the
$\left( 1,1\right) $ mode (see in Fig.\ref{simu}(c)) in the central
region of the condensate, leading to a greater overlap for (1,-1)
with $\left( 0,0\right) $ source mode. This makes the coupling
between the $\left( 1,-1\right) $ and $\left( 0,0\right) $ modes\
stronger than that for the $\left( 1,1\right) $ and $\left(
0,0\right) $ modes. Therefore, for long pulse durations, there are
more atoms being\ scattered to ($1,-1$) mode by the $e_{+}$ light,
leading to a higher atomic population in the ($1,-1$) side-mode.
This phenomenon has been observed in the experiment as shown in
Fig.~\ref{simu}(b).

As the duration of the pulse field further increases, high order
side-modes such as the $\left( 2,0\right) $ mode begin to
participate in the matter-wave superradiant process, as shown in
\ref{simu}(d). At $t =120\mu s$, there has been a considerable
number of atoms that are excited to the $\left( 2,0\right) $ mode.
The pump laser, together with the light fields in all the end-fire
modes, couples the different atomic side-modes with each other. This
makes the atomic superradiance a process rather like the atomic
Bragg diffraction. Besides the fact that the $\left( 1,-1\right) $
mode is stronger than the $\left( 1,1\right) $ mode, the atomic
population in each side-mode begins to oscillate in the time domain.

\section{Conclusion}
In this paper, we have demonstrated experimentally and investigated
asymmetric matter-wave superradiant scattering of a distorted BEC by
measuring the dynamics of atoms in the two dominantly occupied
first-order side-modes. It is found that, even a tiny atomic density
disturbance would lead to prominent asymmetric atomic distributions
among the different superradiant side-modes in the long pulse
regime. The matter-wave side-modes are amplified in an unexpected
way because of the abnormal magnification of the end-fire mode in
diluter part of the BEC resulted from the mode competitions in the
central part of the condensate. By means of superradiance, we have
successfully signalized the symmetry of the atomic distribution of
the condensate and finally measured it in the experiment. Our
experiment is stable against the non-stationary disturbance, and
presents a new way to detect the geometric symmetries of the density
distribution of BEC.

\section*{Acknowledgments}
We appreciate H. W. Xiong, B. Wu for helpful discussions. This work
is supported by the National Fundamental Research Program of China
under Grant No. 2011CB921501, the National Natural Science
Foundation of China under Grant No. 61027016, No.61078026 and
No.10934010, the RFDP No.20120001110091 and NCET-10-0209.

\appendix{}
\section{The influence from the atomic acceleration}
In this appendix, we are going to prove that, the acceleration of
the condensate will not add asymmetry to the matter-wave
superradiance interference patterns, which is usually not true for
the ordinary matter-wave interference experiment. This will make our
experimental technique a rather stable method in detecting the
geometric symmetry of BEC or in other potential applications against
dynamical disturbance.

\begin{figure}[!h]
\begin{center}
\includegraphics[width=8cm]{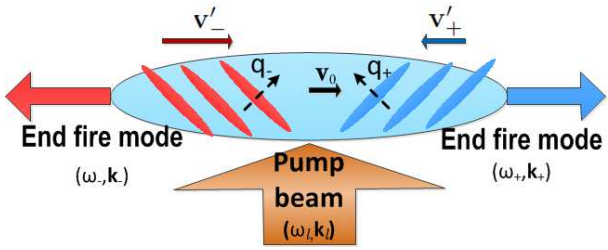}
\end{center}
\caption{Schematic of matter-wave superradiance of a condensate
moving at a
velocity $v_{0}$. There are two moving optical lattices (with wave vector $%
\mathbf{q}_{\pm }$) formed by the superposition the pump pulse and
the light
field in each of the end-fire modes, moving with velocity $v^{\prime}_{\pm }$ along $%
z $ direction, respectively.} \label{opt}
\end{figure}

In the experiment, the condensate has been accelerated by the
unbalanced magnetic trap to move at a velocity $v_{0}$ in the
positive $z$ direction. However, by changing the reference frame to
the one moving with the condensate, it can be find that\
Eq.~(\ref{Semi}) is invariant under such a Galilean transformation.
In fact, the pump laser together with each of the generated end-fire
modes (with wave vector $\mathbf{k}_{\pm }$ and frequency
$\omega _{\pm }$) creates a pair of moving optical lattices with intensity $%
I_{\pm}(\mathbf{r},t)\propto cos(\mathbf{q}_{\pm }\cdot
\mathbf{r}-\Delta \omega
_{\pm }t)$ as shown in Fig.~\ref{opt}, in which $\mathbf{q}_{\pm }=\mathbf{k}%
_{l}-\mathbf{k}_{\pm }$, $\Delta \omega _{\pm }=\omega _{l}-\omega
_{\pm }$. In stationary case, i.e., $v_{0}=0$, we have $\omega
_{+}=\omega _{-}$, and the speeds of these optical lattices are both
$\sqrt{2}\omega _{r}/k$ along their own wave vectors. That is to
say, these two generated optical lattices
move against each other in $\mathbf{\hat{z}}$ direction at a$\ $speed $%
2\omega _{r}/k$.

When the condensate has an initial velocity $v_{0}=2\omega _{r}m^{\prime }/k$%
, where $m^{\prime }$ can be fractional, the frequency of each of
the scattered light will have a Doppler shift, leading to the
modification,
\begin{equation}
\Delta \omega _{\pm }\rightarrow \Delta \omega _{\pm }^{\prime
}=2\omega _{r}(1\pm m^{\prime }),  \label{doppler}
\end{equation}%
and resulting in a new sets of lattices with modified\ speed $v_{\pm
}^{\prime }=2\omega _{r}(\mp 1+m^{\prime })$ in $\mathbf{\hat{z}}$
direction, i.e., compared to the stationary case, the speed of the
generated optical lattices are both increased by $2\omega
_{r}m^{\prime }/k$, which is exactly the same as\ $v_{0}$. In other
words, the matter-wave superradiance of a moving condensate is
equivalent with performing the superradiant experiment of a
stationary BEC in a moving laboratory. Therefore, the atomic
acceleration will induce\ no asymmetries in the atomic dynamics in
the matter-wave superradiant\ processes.

Even though, the atomic acceleration appends a phase factor$\
e^{i(k^{\prime
}x-\omega ^{\prime }t)}$ (where, $k^{\prime }=\frac{Mv_{0}}{\hbar }$, $%
\omega ^{\prime }=\frac{Mv_{0}^{2}}{2\hbar }$) to the wave function
of the condensate, this space and time dependent factor can be
canceled by a unitary transformation. Therefore, it will not
influence the symmetry of superradiant scattering in time domain,
even though it seems that the directed acceleration of the
condensate breaks the spacial inversion symmetry of the system and
may lead to asymmetric dynamics in the matter-wave superradiant
process. Thus any phase disturbance induced from the acceleration of
the condensate by instability of the experimental setup will not
influence the symmetry of the measured matter-wave superradiance
interference patterns. This makes our experiment a rather stable
technique for investigating the matter-wave superradiance against
non-stationary disturbances.

\end{document}